# Optimal Power Control in Decentralized Gaussian Multiple Access Channels

Kamal Singh

***Abstract*—We consider the decentralized power optimization problem for Gaussian fast-fading multiple access channel (MAC) so that the average sum-throughput is maximized. In our MAC setup, each transmitter has access to only its own fading coefficient or channel state information (CSI) while the receiver has full CSI available at all instants. Unlike centralized MAC (full CSIT MAC) where the optimal powers are known explicitly, the analytical solution for optimal decentralized powers does not seem feasible. In this letter, we specialize alternating-maximization (AM) method for numerically computing the optimal powers and ergodic capacity of the decentralized MAC for general fading statistics and average power constraints. For illustration, we apply our AM method to compute the capacity of MAC channels with fading distributions such as Rayleigh, Rician etc.**

## I. Introduction

The multiple access channel (MAC) is a commonly used model to represent communication scenario where multiple senders communicate to a common receiver, such as the uplink channel of a mobile cellular network. The availability of the CSI at the transmitters and receiver has a significant impact on the achievable throughput rates of the fading MAC channels. Under full CSI at the receiver and partial CSI at the transmitters, the ergodic capacity region of a MAC with additive white Gaussian noise (AWGN) and fast fading is completely characterized by the optimal power control schemes [1]. An intuitive justification for this property is that in a fast fading scenario, each codeword experiences all possible fading realizations and thus, any rate close to ergodic capacity can be achieved by choosing to transmit all codewords with the same rate and optimal power strategies [2]. More precisely, Gaussian codebooks with optimal power control achieves the ergodic capacity region, see Figure 1. Depending upon the availability of channel state information (CSI) at the transmitters and receiver, the optimal power control strategy varies. In this letter, we consider a fading MAC where each transmitter knows only its own fading coefficients and the receiver has full CSI. Further, we assume independent fading statistics, with average power constraints not necessary identical, across users. Also, we assume a fast fading model where the channel varies IID (independent and identically distributed) in time.

The power and ergodic capacity problem of the decentralized fading Gaussian MAC is a long-standing open problem posed by Shamai and Telatar in [4] suggesting that analytical solution is not feasible. As an alternative, near closed-form lower bounds on ergodic capacity are derived for the *identical user*[1] MAC channel using a simple heuristic ON-OFF power scheme which further improve with number of MAC users [4]. In [5], structural properties of the optimals are derived to design suitable power schemes to further raise these near closed-form lower bounds. More recently in [6], tight numerical bounds (upper and lower both) to ergodic capacity are obtained for the decentralized MAC for *identical users* setting.

This letter is a first attempt to solve numerically the optimal powers and ergodic capacity problem of the decentralized fading MAC for general fading statistics and average power constraints. Thus, in our study, the identical users MAC is a special case. Our main contribution is a simple alternate maximization based numerical algorithm for optimal decentralized powers and capacity where each of the partial maximizations is solved utilizing the monotone structure of the optimals.

The organization of the letter is as follows. Section II details the system model and the optimization problem to be solved. In Section III, the computational algorithm for the optimal powers based on alternating maximization approach is explained followed by proofs of convergence and optimality. Numerical results for the decentralized fading MAC are presented in section IV. Section V concludes the letter.

## II. System Model

Consider a $L$-user Gaussian fading MAC whose output is given by

$$Y = \sum_{i=1}^{L} H_i X_i + Z,$$

where user-$i$ transmit symbol $X_i$ undergoes flat fading denoted by multiplicative coefficient $H_i$. The additive noise $Z$ is a normalized AWGN process independent of $X_i$ and $H_i$. The fading processes $H_i$ are assumed to be independent across users and varies IID in time. In our decentralized model, the fading coefficients $H_i$ are known only to the respective transmitters at all instants. The receiver has access to the full CSI vector $(H_1, H_2, \cdots, H_L)$. We also assume that the fading distributions are known *a priori* to all the transmitters and the receiver. The $i$-th transmitter, using the available channel state information $h_i$ (current realization of $H_i$), selects transmit power of $P_i(h_i)$, see Figure 1. For convenience, with a slight abuse of notation, we will use $P_i$ to denote the power control of the $i$-th user. For a chosen set of power schemes denoted by $(P_1, \cdots, P_L)$, the average ergodic sum-rate $R \triangleq \mathbb{E} \sum_{i=1}^{L} R_i$ given by

$$R(P_1, \cdots, P_L) = \mathbb{E} \log\left(1 + \sum_{i=1}^{L} |H_i|^2 P_i(H_i)\right), \quad (1)$$

is achieved by employing *successive cancellation* decoding at the receiver [3, Chapter 4]. Since $H_i$ is known at the respective

Kamal Singh is with the Department of Electrical Engineering at IIT Bombay, Mumbai, INDIA-400076 (e-mail: {kamalsingh}@ee.iitb.ac.in).

[1]Fading distributions and power constraints are identical across users.

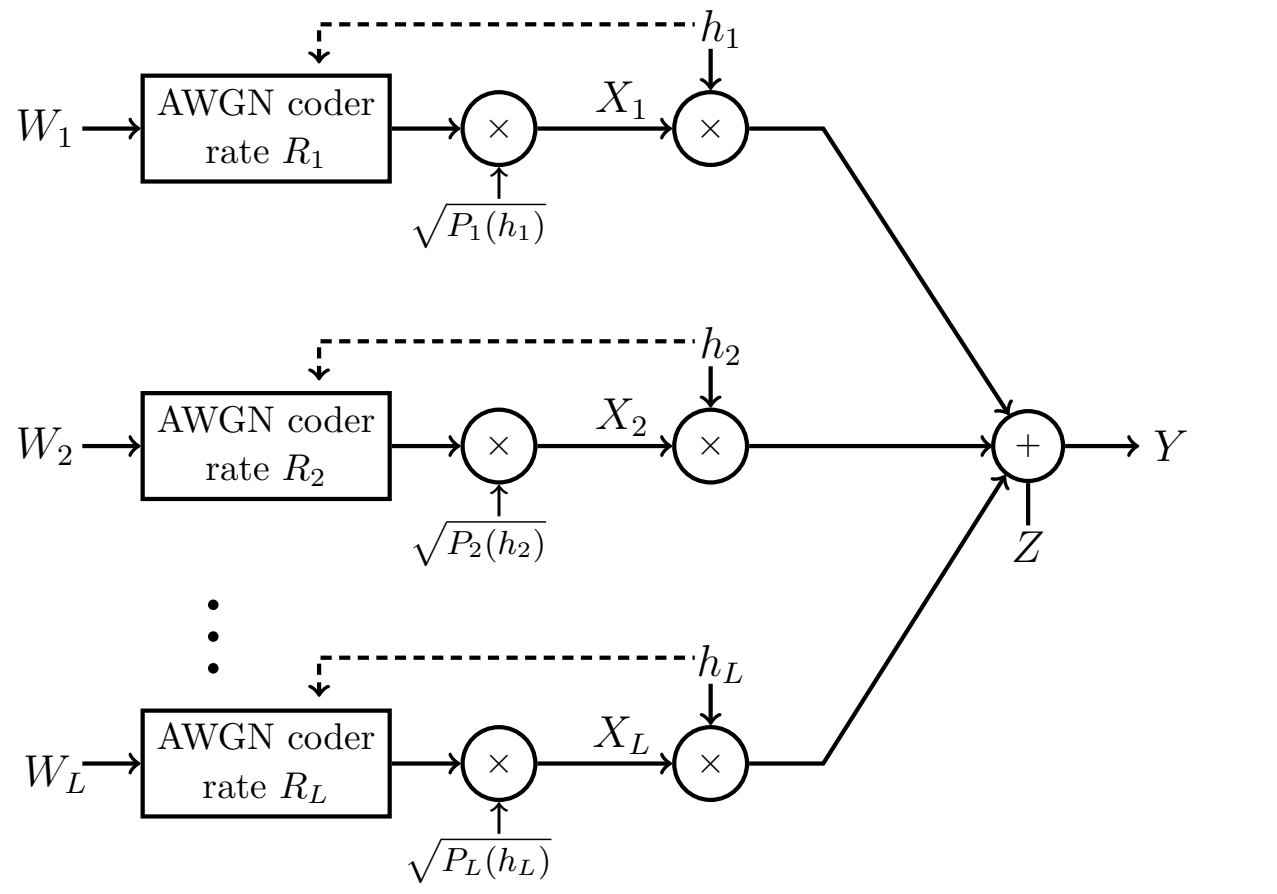

Fig. 1: Power control in Decentralized Gaussian fading MAC.

transmitter and receiver, the sum-rate in (1) depends only on the fading magnitudes. Thus, we can replace $|H_i|^2$ by $V_i$ and write $P_i(H_i)$ as $P_i(V_i)$, $1 \le i \le L$. Our objective is to maximize the sum-rate $R$ over the set of power control schemes $P_i(V_i)$, $1 \le i \le L$, satisfying power constraints associated with the transmitters.

**Definition 1.** *The **ergodic sum-capacity** $C_{sum}$ is the maximum average sum-rate achievable [1], i.e.*

$$C_{sum} = \max_{(P_1,\dots,P_L)\in\mathcal{P}} \mathbb{E}\log\left(1+\sum_{i=1}^{L} V_i P_i(V_i)\right), \quad (2)$$

*where the maximization is over the set $\mathcal{P}$ defined as collection of all power strategies satisfying $\mathbb{E}\,P_i(V_i) \le P_i^{\text{avg}}$, $1 \le i \le L$.*

**Remark 2.** *Notice that the average power constraints are linear and the objective function $R$ is concave in powers $P_i$, $1 \le i \le L$. Furthermore, $R$ is continuous and has continuous partial derivatives. Also, it can be easily deduced that the set $\mathcal{P}$ is a non-empty convex compact set.*

We will use boldface letters to denote vectors. For example, $\mathbf{d}$ is a vector with element $d_i$ at position-$i$ where $d_i$ can be either a scalar or a function. We also use the notation $\mathbf{d}_{\hat{j}}$ to represent a vector containing all elements of $\mathbf{d}$ excluding element $d_j$. The joint distribution (CDF) is denoted by $\Psi$, where $\Psi_i$ denotes the marginal CDF of fading of user-$i$.

## III. Optimal Power Control

Since (2) is a convex program with a strictly feasible point, KKT conditions are necessary and sufficient condition for the optimal powers. We obtain the cost function

$$\mathcal{L} \triangleq \int \log\left(1+\sum_{i=1}^{L} v_i P_i(v_i)\right) d\Psi(\mathbf{v}) - \sum_{i=1}^{L}\lambda_i \int P_i(v_i)\,d\Psi_i(v_i)$$

where the constants $\lambda_i$, $1 \le i \le L$ are Lagrange multipliers for each of the power constraints. The derivatives with respect to the power allocation functions has to be zero for optimality, whenever non-zero power is allocated. Thus,

$$\int \frac{d\Psi(\mathbf{v}_{\hat{j}})}{1+v_jP_j(v_j)+\sum_{i=1,i\neq j}^{L} v_iP_i(v_i)} = \frac{\lambda_j}{v_j},\ 1 \le j \le L. \quad (3)$$

The analytical solution of (3) for the optimal powers is considered not feasible [4]. Next, we identify a key structural property of optimal decentralized powers that enables the computation of optimal powers and ergodic capacity numerically.

**Theorem 3.** *The optimal power $P_j^*(v_j)$, whenever non-zero, must be a monotonically increasing function of $v_j$, $1 \le j \le L$.*

*Proof:* W.l.o.g consider the optimal power scheme say $P_k^*$ of user-$k$. Using (3), we have

$$v_k \int \frac{d\Psi(\mathbf{v}_{\hat{k}})}{1+v_kP_k^*(v_k)+\sum_{i=1,i\neq k}^{L} v_iP_i^*(v_i)} = \lambda_k, \quad (4)$$

whenever $P_k^* > 0$. Furthermore, consider any two values of the fading variable $v_k$ say $\beta > \alpha$ such that positive powers are allocated. Thus, we have

$$\beta \int \frac{d\Psi(y)}{1+\beta P_k^*(\beta)+y} = \alpha \int \frac{d\Psi(y)}{1+\alpha P_k^*(\alpha)+y}, \text{ i.e.}$$

$$\int \left(\frac{\beta}{1+\beta P_k^*(\beta)+y} - \frac{\alpha}{1+\alpha P_k^*(\alpha)+y}\right) d\Psi(y) = 0. \quad (5)$$

where $\sum_{i=1,i\neq k}^{L} v_iP_i^*(v_i)$ is replaced by $y$ for ease of representation. Rewriting the integrand in (5), we get

$$\int \left(\frac{(\beta-\alpha)(1+y)+\alpha\beta(P_k^*(\alpha)-P_k^*(\beta))}{(1+\alpha P_k^*(\alpha)+y)(1+\beta P_k^*(\beta)+y)}\right) d\Psi(y) = 0\cdot \quad (6)$$

The integrand above is strictly positive for $P_k^*(\beta) \le P_k^*(\alpha)$, thus violating (6). Therefore, $P_k^*(\beta) > P_k^*(\alpha)$. This completes the proof. ■

**Remark 4.** *Though we presented proof of monotonicity of the optimal $P_k^*$ assuming that all remaining powers are optimal, this is also true for the partially optimal say $\hat{P}_k^*$ for any set of feasible powers for the remaining users and can be proved using reasoning similar to that in the proof of the Theorem 3.*

Let us denote the integral on the LHS in (3) by $f_j(v_jP_j(v_j))$.

**Corollary 5.** *For optimal $P_j(v_j)$, $f_j(v_jP_j(v_j))$ is monotonically decreasing function of $v_j$.*

*Proof:* For the optimal case, by Theorem 3, $v_jP_j(v_j)$ increases monotonically in $v_j$ and the corollary follows. ■

### A. AM Algorithm

We now develop a simple numerical algorithm to solve the joint optimization problem in (2) in terms of partial optimizations using the principle of alternating maximization (AM). As we will later prove, the partial maximization can be solved numerically using the monotone structure of the optimal. The convergence and the optimality proofs are established in the next sub-section.

The alternating maximization (AM) method maximizes $R$ w.r.t. each power scheme sequentially. To identify this, we denote the $i$-th user power by $P_i^{(n)}$. The computational algorithm is parameterized in terms of $\lambda_i$, $1 \le i \le L$.

**Algorithm** Optimal powers for decentralized MAC

**Initialization:** Initialize $\lambda_j$, $1 \le j \le L$, small step-size $\delta$, approximation error tolerance $\epsilon$. $P_j^{(0)}$, $1 \le j \le L$ denote arbitrarily initialized powers obeying constraints. Set $n = 1$.
**Repeat**
**For** $j = 1$ to $L$
(a) Compute the partial maximization

$$P_j^{(n)} = \arg\max_{P_j} R(P_1^{(n)}, \cdots, P_{j-1}^{(n)}, P_j, P_{j+1}^{(n-1)}, \cdots, P_K^{(n-1)}),$$

using the formula: $P_j^{(n)}(v_j) = \frac{1}{v_j} f_j^{-1}\left(\frac{\lambda_j}{v_j}\right)$.

(b) Find $\bar{P}_j^{\text{avg}} = \int P_j^{(n)}(v_j) d\Psi_j(v_j)$.

(c) If $\begin{cases} \left(P_j^{\text{avg}} - \bar{P}_j^{\text{avg}}\right) > \epsilon, \text{ then } \lambda_j = \lambda_j + \delta; \text{ goto step (a)} \\ \left(P_j^{\text{avg}} - \bar{P}_j^{\text{avg}}\right) < -\epsilon, \text{ then } \lambda_j = \lambda_j - \delta; \text{ goto step (a)} \end{cases}$

**End**
$n = n + 1$
**Until all the power constraints converge.**

Notice that the algorithm outputs power schemes in the order $P_1^{(0)}, P_2^{(0)}, \cdots, P_K^{(0)}, P_1^{(1)}, P_2^{(1)}, \cdots, P_K^{(1)}, P_1^{(2)}, P_2^{(2)}, \cdots, P_K^{(2)}, \cdots$, where each power in the sequence is the partial maximizer in step (a) with the previously available powers fixed for the remaining users. In the following corollary, we justify the procedure for the partial maximization in step (a).

**Corollary 6.** *The partial maximizer $P_j^{(n)}$ is solved by*

$$P_j^{(n)}(v_j) = \frac{1}{v_j} f_j^{-1}\left(\frac{\lambda_j}{v_j}\right), \ 1 \leq j \leq L, \tag{7}$$

*where $f_j^{-1}(\cdot)$ is the inverse mapping of the $f_j(\cdot)$ function.*

*Proof:* Since for optimal $P_j(v_j)$, $f_j(v_j P_j(v_j))$ is strictly decreasing with $v_j$, there is a one-to-one correspondence between the RHS and LHS of (3) for every $v_j$. Hence $f_j^{-1}(\cdot)$, inverse mapping of $f_j(\cdot)$, exists and the corollary follows. ■

Precisely speaking, for every $v_j$, we compare the computed integral $f_j(v_j P_j(v_j))$ with $\frac{\lambda_j}{v_j}$ for different values of $P_j(v_j)$ until the two values agree to the desired accuracy. This is done using bisection method (linear convergence, rate 1/2) to solve (3) for $P_j(v_j)$. The algorithm repeats until convergence (step (c)) with stop condition as $|P_j^{\text{avg}} - \bar{P}_j^{\text{avg}}| < \epsilon$, $1 \leq j \leq L$.

### B. Convergence and Optimality

In general, alternating optimizations need not converge. We now show that AM algorithm always converges to the global optimal of (2). The proof follows a set of arguments similar to convergence and optimality proofs of AM optimization for convex objective function and convex constraints presented in [8, Chapter 9] with appropriate modifications.

We define term "iteration" to indicate updating of all $L$-user's power exactly once from previously updated $L$-user's powers set. Let $\mathbf{P}^{(n)} := (P_1^{(n)}, P_2^{(n)}, \cdots, P_L^{(n)})$ denote the updated all powers set after iteration-$n$.

**Lemma 7.** *There exists a constant $R^*$ such that*

$$R(\mathbf{P}^{(n)}) \to R^*$$

*Proof:* We describe the proof for $L = 2$ user case. The extension to higher $L$ is straightforward. For every run of the algorithm, the ergodic sum-rate improves i.e.

$$R(P_1^{(n-1)}, P_2^{(n-1)}) \leq R(P_1^{(n)}, P_2^{(n-1)}) \leq R(P_1^{(n)}, P_2^{(n)}).$$

In short, $R(\mathbf{P}^{(n)}) \geq R(\mathbf{P}^{(n-1)})$ holds for all $n$. Since the rate sequence $R(\mathbf{P}^{(n)})$ is non-decreasing and bounded from above, it must converge i.e. $R(\mathbf{P}^{(n)}) \to R^*$ for some $R^* \leq C_{sum}$. ■

In our main Theorem 9, we show that the AM algorithm attains the global optimum irrespective of the chosen starting or initializing conditions. Towards this end, we define

$$\Delta R(\mathbf{P}) = R(\mathbf{P}_{next}) - R(\mathbf{P}),$$

where $\mathbf{P}_{next}$ is the updated powers set after an iteration of the AM algorithm using $\mathbf{P}$ as previous powers set. Thus, $\Delta R(\mathbf{P})$ is the increment in $R(\mathbf{P})$ after an iteration of the algorithm.

**Corollary 8.** *If $\Delta R(\mathbf{P}) = 0$ for any $\mathbf{P} \in \mathcal{P}$, then $\mathbf{P}_{next} = \mathbf{P}$.*

*Proof:* The condition $\Delta R(\mathbf{P}) = 0$ implies there is no increment in the sum-rate in each of the partial optimizations. This, in turn, implies $P_{next,j} = P_j$, $1 \leq j \leq L$ due to uniqueness of solutions of partial maximizations i.e. each $P_j$ is the partial optimizer when remaining powers are fixed. ■

**Theorem 9.** *The AM Algorithm converges to global optimum i.e.*

$$R^* = C_{sum}.$$

*Proof:* The proof consists of two parts: (1) showing that if $R(\mathbf{P}^{(n)}) < C_{sum}$ for any power $\mathbf{P}^{(n)}$, then $R(\mathbf{P}^{(n+1)}) > R(\mathbf{P}^{(n)})$ i.e. the algorithm does not get trapped if $R(\mathbf{P}^{(n)}) < C_{sum}$, and (2) showing that $R(\mathbf{P}^{(n)})$ necessarily converges to $C_{sum}$.

**Part (1):** If $R(\mathbf{P}) < C_{sum}$ for any $\mathbf{P} \in \mathcal{P}$, then

$$R(\mathbf{P}_{next}) > R(\mathbf{P}) \ \text{ or } \ \Delta R(\mathbf{P}) > 0.$$

The proof is by contradiction. Consider any $\mathbf{P} \in \mathcal{P}$ such that $R(\mathbf{P}) < C_{sum}$. Assume $\Delta R(\mathbf{P}) = 0$. Since $R(\mathbf{P}) < C_{sum}$, there exists $\mathbf{Q} \in \mathcal{P}$ such that $R(\mathbf{P}) < R(\mathbf{Q})$. Consider the direction from $\mathbf{P}$ to $\mathbf{Q}$, we see that

$$\mathbf{Q} - \mathbf{P} = [Q_1 - P_1, 0, 0, \cdots, 0] + [0, Q_2 - P_2, 0, \cdots, 0] + \cdots + [0, \cdots, 0, 0, Q_L - P_L].$$

Rewriting the above in unit vector form, we get

$$\vec{\mathbf{u}} = \alpha_1 \vec{\mathbf{u}}_1 + \alpha_2 \vec{\mathbf{u}}_2 + \cdots + \alpha_L \vec{\mathbf{u}}_L,$$

where $\vec{\mathbf{u}}$ is unit vector along $\mathbf{Q} - \mathbf{P}$, $\vec{\mathbf{u}}_1$ along $[Q_1 - P_1, 0, 0, \cdots, 0]$, $\vec{\mathbf{u}}_2$ along $[0, Q_2 - P_2, 0, \cdots, 0]$ etc. and $\alpha_i = \|Q_i - P_i\| / \|\mathbf{Q} - \mathbf{P}\|$, $1 \leq i \leq L$ are the scalars. The rate of change of $R(\mathbf{P})$ in the direction from $\mathbf{P}$ to $\mathbf{Q}$ is given by

$$\begin{aligned} \nabla R(\mathbf{P}) \cdot \vec{\mathbf{u}} &= \nabla R(\mathbf{P}) \cdot (\alpha_1 \vec{\mathbf{u}}_1 + \alpha_2 \vec{\mathbf{u}}_2 + \cdots + \alpha_L \vec{\mathbf{u}}_L), \\ &= \alpha_1 \nabla R(\mathbf{P}) \cdot \vec{\mathbf{u}}_1 + \alpha_2 \nabla R(\mathbf{P}) \cdot \vec{\mathbf{u}}_2 + \cdots + \alpha_L \nabla R(\mathbf{P}) \cdot \vec{\mathbf{u}}_L. \end{aligned}$$

Consider the direction $\vec{\mathbf{u}}_1$:

$$[Q_1 - P_1, 0, 0, \cdots, 0] = [Q_1, P_2, P_3, \cdots, P_L] - [P_1, P_2, P_3, \cdots, P_L].$$

Corollary 8 implies $P_1$ maximizes sum-rate $R$ for the fixed set of remaining powers $\{P_2, P_3, \cdots, P_L\}$. Hence $\nabla R(\mathbf{P}) \cdot \vec{\mathbf{u}}_1 = 0$. By similar arguments, $\nabla R(\mathbf{P}) \cdot \vec{\mathbf{u}}_i = 0$, $1 \leq i \leq L$ holds. Thus,

$$\nabla R(\mathbf{P}) \cdot \vec{\mathbf{u}} = \sum_{i=1}^{L} \alpha_i \nabla R(\mathbf{P}) \cdot \vec{\mathbf{u}}_i = 0. \tag{8}$$

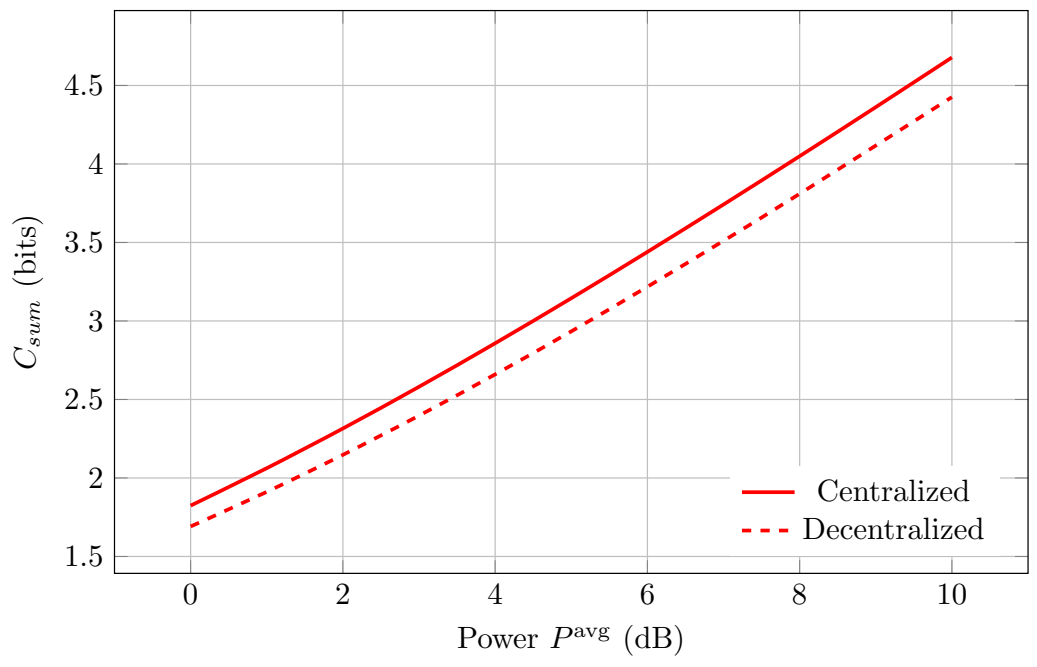

(a)

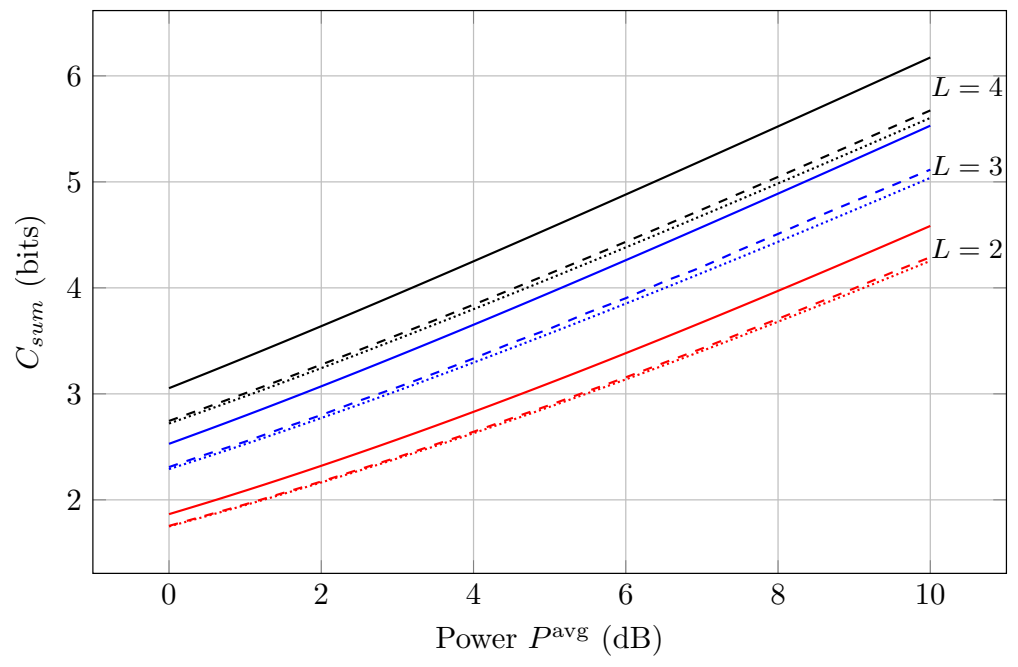

(b)

Fig. 2: (a) Capacity result for $L = 2$ *non-identical users* decentralized MAC with normalized Rayleigh and Rician fading i.e. $d\Psi_1(h_1) = 2h_1 e^{-h_1^2} dh_1$, $d\Psi_2(h_2) = 2h_2(K+1)e^{-h_2^2(K+1)-K} I_0(2h_2\sqrt{K(K+1)})dh_2$ and $\mathbb{E}[h_1^2] = \mathbb{E}[h_2^2] = 1$. Solid curve is capacity of the centralized MAC [7]. (b) Capacity results for *L-identical users* decentralized MAC with all fadings Rayleigh distributed: Dashed curves are the desired capacity of decentralized MAC, dotted curves are near closed-form lower bounds to decentralized MAC capacity obtained in [5] and solid curves represent capacity of centralized MAC [7].

Since sum-rate $R$ is concave, it satisfies

$$R(\mathbf{Q}) \leq R(\mathbf{P}) + \nabla R(\mathbf{P}) \cdot \|\mathbf{Q} - \mathbf{P}\| \, \vec{\mathbf{u}}, \ \forall \, \mathbf{P} \in \mathcal{P}, \ \forall \, \mathbf{Q} \in \mathcal{P},$$

which, with (8), implies $R(\mathbf{Q}) \leq R(\mathbf{P})$ which is a contradiction. Therefore, $\Delta R(\mathbf{P}) > 0$.

**Part (2):** Convergence to the optimal is not established since the algorithm may produce increments $\Delta R(\mathbf{P}) > 0$ arbitrarily small. We now prove that the sum-rate sequence indeed converges to $C_{sum}$. To this end, we recall, from Lemma 7, that the sum-rate $R(\mathbf{P}^{(n)})$ converges to say $R^*$. Thus, for any $\delta > 0$ and for all $n$ sufficiently large, we have

$$R^* - \delta \leq R(\mathbf{P}^{(n)}) \leq R^*. \tag{9}$$

Let
$$\mu = \min_{\widetilde{\mathcal{P}}} \Delta R(\mathbf{P}),$$

where $\widetilde{\mathcal{P}} = \{\mathbf{P} \in \mathcal{P} : R^* - \delta \leq R(\mathbf{P}) \leq R^*\}$. Recall that $R(\mathbf{P})$ is continuous and has continuous partial derivatives. Thus, $\Delta R(\mathbf{P})$ is also continuous. Since $\widetilde{\mathcal{P}}$ is inverse image of a closed interval under continuous $R(\mathbf{P})$ and $\mathcal{P}$ is compact, we conclude that the subset $\widetilde{\mathcal{P}}$ is also compact. Thus, $\mu$ exists.

If $R^* < C_{sum}$, then $\Delta R(\mathbf{P}) > 0$ for all power schemes in $\widetilde{\mathcal{P}}$ (Part (1)) and hence $\mu > 0$. Since $R(\mathbf{P}^{(n)})$ satisfies (9), $\mathbf{P}^{(n)} \in \mathcal{P}_1$. Therefore, $\Delta R(\mathbf{P}^{(n)}) \geq \mu$ holds for all sufficiently large $n$. Since $\mu > 0$, this implies $\Delta R(\mathbf{P}^{(n)}) > 0$ for all sufficiently large $n$ suggesting that $R(\mathbf{P}^{(n)})$ eventually exceeds $R^*$, which is a contradiction. Therefore, $R(\mathbf{P}^{(n)}) \to C_{sum}$. ■

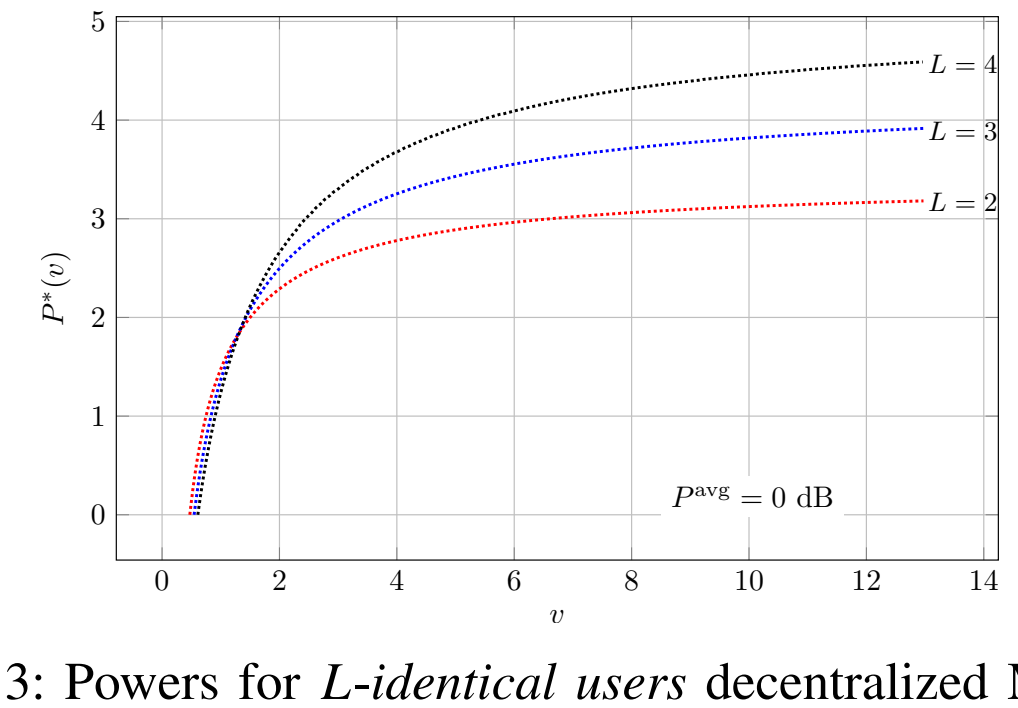

Fig. 3: Powers for *L-identical users* decentralized MAC.

## IV. Numerical results

We demonstrate the utility of the proposed algorithm for Gaussian MAC with independent fadings across users and average power constraints assumed identical for simplicity. Figures 2 and 3 illustrate the ergodic capacity computed for the non-identical & identical users MAC and the optimal powers for the identical users MAC respectively. Here, we do not pursue detailed convergence-rate and complexity analysis of the AM algorithm due to lack of space but empirical estimates suggest that its convergence rate slows down significantly by several orders of magnitude with number of MAC users.

## V. Conclusions

We demonstrated, for the first time, the numerical approach based on alternating optimization principle to solve the decentralized powers and ergodic capacity of Gaussian MAC for the general fading statistics and power constraints. The proposed algorithm is simple to implement but the computational complexity increases with MAC users, thus rendering it useful for MAC with moderate number of users. As future work, we look forward to explore strategies to improve convergence rates of our iterative AM algorithm.